\documentclass[12pt,a4paper,english,amssymb,nofootinbib,superscriptaddress]{revtex4-2}
\usepackage{lmodern}
\usepackage{lmodern}
\usepackage[T1]{fontenc}
\usepackage[latin9]{inputenc}
\usepackage{color}
\usepackage{babel}
\usepackage{verbatim}
\usepackage{amsmath}
\usepackage{amssymb}
\usepackage{esint}
\usepackage[unicode=true,pdfusetitle,
 bookmarks=true,bookmarksnumbered=false,bookmarksopen=false,
 breaklinks=false,pdfborder={0 0 1},backref=false,colorlinks=true]
 {hyperref}
\hypersetup{
 linkcolor=blue,citecolor=blue}

\makeatletter

\pdfpageheight\paperheight
\pdfpagewidth\paperwidth

\@ifundefined{textcolor}{}
{%
 \definecolor{BLACK}{gray}{0}
 \definecolor{WHITE}{gray}{1}
 \definecolor{RED}{rgb}{1,0,0}
 \definecolor{GREEN}{rgb}{0,1,0}
 \definecolor{BLUE}{rgb}{0,0,1}
 \definecolor{CYAN}{cmyk}{1,0,0,0}
 \definecolor{MAGENTA}{cmyk}{0,1,0,0}
 \definecolor{YELLOW}{cmyk}{0,0,1,0}
}

\usepackage{babel}
\usepackage{babel}
\usepackage{babel}
\usepackage{babel}
\usepackage{babel}
\usepackage{babel}
\@ifundefined{definecolor}{color}{}
\usepackage{babel}

\usepackage{latexsym}\usepackage{bm}

\makeatother

\begin{document}

\title{Kerr-Vaidya type radiating black holes in semi-classical gravity with conformal anomaly
}

\author{Metin G{\"u}rses}
\email{gurses@fen.bilkent.edu.tr}

\selectlanguage{english}%

\affiliation{{\small{}Department of Mathematics, Faculty of Sciences}\\
 {\small{}Bilkent University, 06800 Ankara, Turkey}}

\author{Bayram Tekin}
\email{btekin@metu.edu.tr}

\selectlanguage{english}%

\affiliation{Department of Physics,\\
 Middle East Technical University, 06800 Ankara, Turkey}
\begin{abstract}
\noindent Static black holes in the conformal anomaly-sourced semi-classical General Relativity in four dimensions were extended to rotating, stationary solutions, recently.  These quantum-corrected black holes show different features compared to the Kerr black hole and need for further extensions. Here we remove the condition of stationarity and find radiating (Kerr-Vaidya-type) solutions in the same theory augmented with a cosmological constant. As long as the coupling constant $\alpha$ of the $A$-type trace anomaly is non-zero, we show that  $i)$ the cosmological constant is bounded from above, i.e $\Lambda \le \frac{3}{4 \alpha}$; $ii)$ static black holes exist but they may not be unique; $iii)$ static black holes do not satisfy the second law of black hole thermodynamics; $iv)$ static black holes may have unstable inner horizons; $v)$ In the nonstationary and axially symmetric case, stability of the event horizon and the second law of thermodynamics black holes are problematic.
\end{abstract}
\maketitle

\section{Introduction}

In the absence of a consistent framework for quantum gravity, a semi-classical approach to General Relativity (GR) (where matter fields are taken to be quantum fields, while the geometry is kept classical)  has borne much fruit since the early 1970s, especially within the context of black holes which amplify quantum effects: Hawking radiation \cite{Hawking} removes the utter dullness in the lives of stationary and static black holes of classical GR and make them evaporate and shrink.  One particular semi-classical approximation is built on the conformal anomaly for which one has the full knowledge of the trace of the expectation value energy-momentum tensor operator within any quantum state describing classically conformally invariant fields. Recently, Fernandes \cite{Fernandes} found stationary and axially-symmetric rotating black hole solutions in GR, without a cosmological constant, but with a source that comes from the trace anomaly induced by the 1-loop effects of the quantum fields within the semi-classical approximation. These solutions (of which uniqueness is not yet known) demonstrate various novel features in contrast to their vacuum GR limits such as the violation of the Kerr bound and the non-symmetric event-horizons. The solutions given in \cite{Fernandes} generalize the earlier static and spherical symmetric black hole solutions in the same theory given by Cai {\it et al.} \cite{Cai1} where an important stumbling block in finding the solutions of the anomaly-sourced theory (to be explained below) was also circumvented. In \cite{Cai2}, a negative cosmological constant was introduced, and the static, spherically symmetric solutions were found and their thermodynamics was studied.

The current state of the trace-anomaly sourced semi-classical GR was nicely described in \cite{Fernandes} and we invite the interested reader to refer to that work, but here let us briefly describe the theory. It is well-known that even if one starts from a classically conformally invariant theory (say a theory of massless fields conformally coupled to gravity in four dimensions), the symmetry does not generally survive quantization (more properly regularization) at a one-loop level  \cite{Duff}.  The Weyl-scaling invariance of the metric is lost, and this shows itself in the non-zero trace of the expectation value of the energy-momentum tensor (in any quantum state). Even without a detailed knowledge of the massless quantum fields coupled to gravity, we know that in four dimensions, the trace anomaly is expressed purely in terms of the curvature invariants of the background spacetime with the metric $g_{\mu \nu}$ as
\begin{equation}
\langle \psi | T|\psi  \rangle := g^{\mu \nu} \langle \psi | T_{ \mu \nu}|\psi  \rangle = \frac{\beta}{2} C_{\mu\nu \sigma\rho} C^{\mu\nu \sigma\rho} - \frac{\alpha}{2} \mathcal{G}, \label{an}
\end{equation}
where  $C_{\mu\nu \sigma\rho}$ are the components of the Weyl tensor in some coordinates, and $\mathcal{G}$ is the Gauss-Bonnet scalar defined in terms of the Riemann, Ricci tensors and the scalar curvature as
\begin{equation}
\mathcal{G} :=  R_{\mu\nu \sigma\rho} R^{\mu\nu \sigma\rho}- 4R_{\mu \nu} R^{\mu \nu} + R^2.
\end{equation}
In (\ref{an}), the constants $\alpha$ and $\beta$ are the only inputs coming from the underlying conformal field theory; and are known explicitly in terms of the number of massless fields \cite{Deser1}. The fact that the right-hand side of (\ref{an}) does not have information about the quantum fields (except their number, just mentioned) is a blessing: one can study the backreaction of these fields (in a semi-classical setting of course) on the geometry they happen to live in. But it is clear that without the full energy-momentum tensor, one can find only the solutions to the trace of the field equations. That trace equation is only a {\it necessary} condition, but not a {\it sufficient} one in general. That means the general solution to the trace of the field equations will involve the correct solution as a subclass, but it probably will be further restricted by the full theory. This is the apparent impasse in considering the anomaly-sourced field equations, and generically there is no known solution yet: one must compute the full tensor $\langle \psi | T_{ \mu \nu}|\psi  \rangle$ which does involve details of the quantum fields.  So even at the semi-classical level, it is very hard to compute the backreaction of the quantum corrections to a generic background metric $g_{\mu \nu}$.

Furthermore, as a second issue, one does need good reasons to calculate such corrections to the classical background. The main question is the following: can there be macroscopic effects of these corrections for example in a black hole geometry where strong gravity amplifies apparently small effects? This second issue was settled in the affirmative in \cite{Mottola}: quantum conformal anomaly can have macroscopic effects. Let us also note that in a recent work \cite{Horo}, extremal black holes were shown to amplify quantum effects, generically, not just the ones coming from the conformal anomaly.
Therefore, there is ample reason to study the backreaction of the conformal anomaly in a black hole background. In the first issue of the full energy-momentum tensor, a tentative but very useful solution is the following: assume some symmetry in the background geometry together with some simplifying assumptions, such as staticity, spherical symmetry, stationarity, {\it etc.} to fix the total energy-momentum tensor.  (For this discussion see the relevant literature in \cite{Fernandes}). In this work, we remove the important assumption of {\it stationarity}, that is we assume that the spacetime does not have a time-like vector field, and this leads to a generalization of the rotating stationary black holes of  Fernandes \cite{Fernandes}, and the static black holes of Cai {\it et al.} \cite{Cai1, Cai2} to dynamical black holes with radiation either emitted by the black hole or absorbed by it. We also include a cosmological constant generalizing the solution in \cite{Cai2}.  As we shall discuss, our solution describes a quantum-corrected version of the spherically symmetric and rotating Vaidya-type radiating solution \cite{Vaidya}.

The layout of the paper is as follows: In Section II, we discuss the non-static spherically symmetric solution, that is the conformal anomaly sourced (radiating, or radiation absorbing) Vaidya metric, In Section III we discuss the rotating version, and we delegate the rather long expression of the full energy-momentum tensor to the Appendix.
\section{ Spherically Symmetric Radiating solution}

We consider the conformal anomaly-sourced cosmological Einstein gravity (in the units $8 \pi G =1 =c$) as our semi-classical field equations
\begin{equation}
R_{\mu \nu} - \frac{1}{2} g_{\mu \nu} R + \Lambda g_{\mu \nu} =  \langle \psi | T_{ \mu \nu}|\psi  \rangle,
\label{ein}
\end{equation}
together with the usual covariant conservation equation $\nabla^\mu \langle \psi | T_{ \mu \nu}|\psi  \rangle=0$ that comes from the requirement that diffeomorphism invariance survives regularization. The field equations are augmented with the trace anomaly equation (\ref{an}). Then the trace of (\ref{ein}) is a single constraint on the geometry of the underlying spacetime:
\begin{equation}
4 \Lambda- R-\frac{\beta}{2} C_{\mu\nu \sigma\rho} C^{\mu\nu \sigma\rho}+\frac{\alpha}{2}\mathcal{G}=0. \label{trace}
\end{equation}
Let us note again that the constants $\alpha, \beta$ contain information about the quantum fields, but we do not need their explicit forms here. We first consider a spherically symmetric, but non-static metric
\begin{equation}
ds^2=-\left (1-2 m(v,r)\right ) dv^2+2 \epsilon dv dr+r^2 \left( d \theta^2+\sin^2\theta d \phi^2 \right), \label{symmetric}
\end{equation}
where $v$ is the retarded/advanced null coordinate. For the special case of $m(v,r)=m(v)/r$, and $\epsilon =  -1$, the metric describes a Vaidya black hole \cite{Vaidya} emitting radiation, while for $\epsilon =  1$ the radiation is absorbed by the black hole; both of these cases are quite relevant as they describe dynamical black holes with Hawking radiation; or light/ultrarelativistic dust accretion, respectively.
Then the Ricci scalar,  Gauss-Bonnet combination, and the square of the Weyl tensor, which are independent of the sign of $\epsilon$, can be found to be
\begin{equation}
R=2 \frac{\left(r^2 m\right)''}{r^2},\quad \quad \mathcal{G}=\frac{8 \left(m^2\right)''}{r^2}, \quad\quad C_{\mu\nu \sigma\rho} C^{\mu\nu \sigma\rho} = \frac{4 r^2}{3} \left (  \left (\frac{m}{r}\right)^{''} \right )^2,
\end{equation}
where $m' = \partial_r m(v,r)$.  Observe that no derivative with respect to the null coordinate appears in these curvature invariants, even though they appear in the curvature components. For the Vaidya black hole case, that is when $m(v,r)=m(v)/r$, one has a null-dust source.
\begin{equation}
G_{\mu \nu} = T_{\mu \nu} = \frac{2 \epsilon}{r^2} \partial_v m(v) \delta_\mu^v \delta_\nu^v, \hskip 1 cm \text{for the Vaidya black hole}.
\end{equation}
On the other hand, generically, for (\ref{symmetric}), one has a non-null energy-momentum tensor.
\begin{equation}
G^{\mu}\,_\nu=-\frac{2}{r^2}\left(
\begin{array}{cccc}
 \partial_r(r m)& 0 & 0 & 0 \\
 -r \partial_v m &  \partial_r(r m) & 0 & 0 \\
 0 & 0 & r \partial_r(r \partial_r m)& 0 \\
 0 & 0 & 0 & r \partial_r(r \partial_r m) \\
\end{array}
\right).
\end{equation}
The trace equation (\ref{trace}) yields a single non-linear ODE:
\begin{equation}
4 \Lambda - 2 \frac{\left(r^2 m\right)''}{r^2} - \frac{2 \beta r^2}{3} \left (  \left (\frac{m}{r}\right)^{''} \right )^2 +4 \alpha \frac{\left(m^2\right)''}{r^2}=0. \label{ode1}
\end{equation}
For $\beta \ne 0$, an exact analytical solution is not available,\footnote{See \cite{Ho} for a recent numerical approach to this problem.} so as in  \cite{Cai1,Cai2,Fernandes}, we set $\beta =0$ (that is the vanishing of the $B$-type anomaly) and consider only the $A$-type anomaly case with $\alpha \ne 0$. Then, (\ref{ode1}) becomes
\begin{equation}
2 \Lambda r^2 - \left(r^2 m\right)'' +2 \alpha \left(m^2\right)''=0,
\end{equation}
which gives a quadratic equation for $m(v,r)$
\begin{equation}
2 \alpha m^2-r^2 m+\frac{\Lambda}{6}r^4=p(v) r+q(v),
\end{equation}
where $p$ and $q$ are arbitrary differentiable functions of the null cordinate $v$.  The general solution, for $\alpha \ne 0$, is
\begin{equation}
m(v,r)=\frac{r^2}{4 \alpha}\pm \frac{r^2}{4 \alpha} \sqrt{1-\frac{4}{3} \Lambda \alpha  +8 \alpha \left( \frac{p(v)}{r^3}+\frac{q(v)}{r^4} \right)}, \label{static_cozum}
\end{equation}
which is a generalization of the Cai {\it et al.} \cite{Cai1} metric to the non-static case.  Note that the {\it plus} branch is a new solution that diverges as $\alpha \rightarrow 0$, while the {\it minus} branch smoothly goes over to the $\alpha =0$ solution, which is
\begin{equation}
m(v,r) = \frac{\Lambda r^2}{6} - \frac{ p(v)}{r} - \frac{q(v)}{r^2}, \hskip 1 cm  \text{for $\alpha$ =0}.
\end{equation}
The reality of $m(v,r)$in  (\ref{static_cozum}) requires that for all $v$ and $r$, one has
\begin{equation}
1-\frac{4}{3} \Lambda \alpha  +8 \alpha \left( \frac{p(v)}{r^3}+\frac{q(v)}{r^4}\right )  \ge 0.
\end{equation}
For example as $r \rightarrow \infty$, one must have $1-\frac{4}{3} \Lambda \alpha \ge 0$, and for $\alpha >0$, this sets an upper bound on the cosmological
constant of the de Sitter space in terms of the anomaly coefficient. Namely, it must satisfy $\Lambda  \leq \frac{3}{ 4 \alpha}$.
Asymptotically, as $r \rightarrow \infty$, the two branches of (\ref{static_cozum}) behave as
\begin{equation}
m(v,r)\rightarrow  \frac{\left(1\pm \sqrt{\mu}\right) r^2}{4 \alpha }+\frac{p(v)}{\sqrt{\mu} r}+\frac{q(v)}{\sqrt{\mu} r^2}+\mathcal{O}\left(\frac{1}{r^3}\right),
\end{equation}
where $\mu=1- \frac{4\Lambda \alpha}{3}$. This asymptotic behavior shows that at a constant $v$ coordinate, the spacetime is asymptotically a Reissner-Nordstrom-de Sitter (or anti-de Sitter) manifold, with the following identifications  of the effective cosmological constant, mass and electric charge:
\begin{equation}
\Lambda_{\text{eff}} =  \frac{3\left(1\pm \sqrt{\mu}\right)}{2 \alpha }, \quad M(v) = \frac{p(v)}{\sqrt{\mu}}, \quad Q^2 = -\frac{2q(v)}{\sqrt{\mu}}.
\end{equation}
This interpretation requires  $p(v) \ge 0$ and $q(v) \le 0$. The asymptotic behavior of the scalar curvature is as follows:
\begin{equation}
R =\frac{6 \left(1 \pm \sqrt{\mu}\right)}{\alpha } \mp\frac{24 \alpha  p(v)^2}{\mu^{3/2} r^6}+\mathcal{O}\left(\frac{1}{r^7}\right).
\end{equation}
Similarly, the asymptotic behavior of the Gauss-Bonnet combination reads as
\begin{equation}
\mathcal{G} =\frac{6 \left(1 \pm \sqrt{\mu}\right)^2}{\alpha^2 } \mp\frac{48  p(v)^2}{\mu^{3/2} r^6}+\mathcal{O}\left(\frac{1}{r^7}\right).
\end{equation}
On the other hand, near $r \rightarrow 0$, the behavior of the scalar curvature is
\begin{equation}
R= \pm \frac{2 \sqrt{2} q(v)}{r^2 \sqrt{\alpha  q(v)}}\pm\frac{3 \sqrt{2} p(v)}{r \sqrt{\alpha  q(v)}}+\frac{3 \left(4\mp\frac{\sqrt{2} p(v)^2 \sqrt{\alpha  q(v)}}{q(v)^2}\right)}{2 \alpha }+\mathcal{O}\left(r\right),
\end{equation}
while the Gauss-Bonnet combination diverges as
\begin{equation}
\mathcal{G}=\pm\frac{4 \sqrt{2} \sqrt{\alpha  q(v)}}{\alpha ^2 r^2}\pm\frac{6 \sqrt{2} p(v) q(v)}{r (\alpha  q(v))^{3/2}}+\left(\frac{6 (1+ \sqrt{\mu})}{\alpha ^2} \mp\frac{3 \sqrt{2} p(v)^2}{(\alpha  q(v))^{3/2}}\right)+\mathcal{O}\left(r\right).
\end{equation}
Observe that both $R$ and $\mathcal{G}$ require $q(v) >0$ near $r=0$ in contrast to the $r \rightarrow \infty $ expansion.

Let us study the event horizon of this metric defined as a null surface $\mathcal{H}(v,r)=$ constant.\footnote{ Note that, in (\ref{symmetric}), the  $2m(v,r)=1$ is not a null surface, it is not the event horizon, but it is the marginally trapped surface or the apparent horizon, see \cite{geometry} for the geometry of the Vaidya spacetime.}
\begin{equation}
g^{\mu \nu}\, \partial_{\mu} \mathcal{H} \partial _{\nu} \mathcal{H}= \mathcal{H}' \Big ( (1-2m) \mathcal{H}'+2 \epsilon \partial_v\mathcal{H} \Big)=0.
\end{equation}
For $\mathcal{H} :=r-r_{H}(v)=0$, the location of the event horizon is given by a nonlinear first-order differential equation
\begin{equation}
1-2m-2 \epsilon \frac{d \,r_{H}}{dv}=0
\end{equation}
which explicitly reads
\begin{equation}
\epsilon \frac{d\, r_{H}}{dv}=\frac{1}{2}-\frac{r_H^2}{4 \alpha}\mp\frac{r_H^2}{4 \alpha} \sqrt{ \mu +8 \alpha \left( \frac{p(v)}{r_H^3}+\frac{q(v)}{r_H^4} \right)}. \label{denk1}
\end{equation}
The exact analytical solution of this non-linear ODE is not available in its full generality. But let us make some remarks on the solutions.
\begin{enumerate}
\item  The theorem on the existence and uniqueness of the first-order differential equations guarantees that the above equation with the appropriate initial condition has unique a  solution provided that   $p(v)$ and $q(v)$ are continuous functions and $ \mu \, r_{H}^4(v) +8 \alpha \left( p(v)\,r_H+q(v) \right) \ne 0$ for all $v \ge 0$.

\item On the other hand, if  $ \mu \, r_{H}^4(v) +8 \alpha \left( p(v)\,r_H+q(v) \right) = 0$ then (\ref{denk1}) may not have unique solutions. As two examples, we have $r_{H}=\sqrt{2 \alpha}$ (which is possible if $p$ and $q$ are constants and one has $q= -p\sqrt{2 \alpha }-\frac{\alpha  \mu }{2}$);  and as a second solution we have
\begin{equation}
r_H(v) = \sqrt{2 \alpha} \tanh \left ( \frac{ \epsilon v +  v_0 }{ 2\sqrt{2 \alpha}}\right ),
\end{equation}
where $v_0$ is an integration constant. This is possible if
\begin{equation}
q(v)=-p(v)\sqrt{2\alpha}\tanh \left(\frac{v \epsilon +v_0}{2 \sqrt{2\alpha}}\right)-\frac{1}{2} \alpha  \mu  \tanh ^4\left(\frac{v \epsilon +v_0}{2 \sqrt{2\alpha}}\right).
\end{equation}
 For different values of $\epsilon$, the positivity of $r_H$ requires different intervals for $v$. When $\epsilon=1$ (the case of the absorption of radiation by the black hole)  then $v  \in [-v_{0}, \infty)$ and when $\epsilon =-1$ (the case of the emission of radiation by the black hole) then $ v \in (-\infty, v_{0}]$.  Both cases are expected since for  $\epsilon=1$, $r_H$ continues to keep getting larger, while for   $\epsilon=-1$, the process must stop at some future $v_0$ as $r_H$ becomes zero.

\item Since the right-hand side of (\ref{denk1}) changes sign in the interval $v \ge 0$, this differential equation implies that the function  $r_{H}$ is decreasing or increasing with respect to $v$ for certain intervals. Hence the horizon area $A=4 \pi r_{H}^2$ exhibits a similar behavior.  This means that the above differential equation does not satisfy the area law of black hole mechanics, or the second law of black hole thermodynamics, i.e.; $\frac{dA}{dv} >0$  is not valid for all $ v \ge 0$ as is expected for dynamical black holes.

\item When the right-hand side of (\ref{denk1}) vanishes, the corresponding solutions are the critical points or the equilibrium solutions of $r_{H}$. These solutions correspond to the static horizons of Cai {\it et al.} \cite{Cai1}. Let's assume that the static horizon is located at $r^{0}_{H}$. Within our formalism, we can check the important question of the linear stability of these equilibrium solutions or the stability of the static horizons. For this purpose, let $r_{H}(v)=r^{0}_{H}+\varepsilon r_{1}(v)$, where $r_{1}$ is a function that satisfies the linearized  form of (\ref{denk1}):
    \begin{equation}
    \frac{dr_{1}}{dv}=w v_{1}, ~~~~w :=-\left[\frac{r^{0}_{H}}{2 \alpha} \mp \frac{1}{\sqrt{\delta}}\,\left(\frac{\mu}{2 \alpha} (r^{0}_{H})^3+8 \alpha p_{0} \right) \right], \label{denk4}
    \end{equation}
\noindent
where $\delta :=\mu (r^{0}_{H})^4+8 \alpha (p_{0} r^{0}_{H}+q_{0})$. In (\ref{denk4}), we considered both $p$ and $q$ near their static values $p_{0}$ and $q_{0}$, respectively. Equation (\ref{denk4}) implies that the static outer horizon (with the plus sign) is stable but the stability of the inner horizon (with the minus sign) depends on the numerical values of the constants $\mu$ and $\alpha$.

\item Thermodynamics of dynamical black holes is a developing subject, and it is not easy to properly define concepts like temperature, and surface gravity even in the case of quasi-equilibrium. Therefore, we shall only note one proposal for computing the surface gravity of the solution we found above. For this, we follow \cite{Fodor} and see \cite{Nielsen} for a nice review of this topic, where other proposals were also discussed. Given a spherically symmetric metric, of the form
\begin{equation}
ds^2 = - A^2(v,r) \Delta(v,r) dv^2+ 2 A(v,r) d v dr + r^2d\Omega_2^2,
\end{equation}
and let
\begin{equation}
\Delta(v,r):= 1 - 2m(v,r).
\end{equation}
Then the surface gravity on the marginally trapped surface (apparent horizon), or the trapping horizon, for which  $\Delta(v,r_H)=0$,  according to \cite{Fodor} is given as
\begin{equation}
\kappa := \frac{A}{4 r m(v,r)} \Big( 1- 2r \partial_r m(v,r)- 2m(v,r) \Big) + \frac{\dot A}{A}, \hskip 2 cm   \dot A = \partial_v A,
\end{equation}
which, for our metric yields
\begin{equation}
\kappa := -\partial_r m(v,r)|_{r_A}  \quad\ \quad \text{evaluated at }\quad  m(v,r_A) = \frac{1}{2},
\label{kappa}
\end{equation}
where $r_A$ is the radius of the apparent horizon which is equivalent to the radius of the event horizon $r_A = 2 m$ in the Schwarzschild black hole case in this coordinate system. For the Schwarzschild black hole case,  $m(v,r) = 1/2 = m/r$, and hence, (\ref{kappa}) yields the expected, constant value $\kappa = 1/(4 m)$ which is proportional to the Hawking temperature. But in our case, $\kappa$ is a non-trivial function of the null coordinate $v$. 

\end{enumerate}

\section{ Radiating Rotating Solutions}

Let us now consider the axially symmetric but non-stationary metric in the coordinates $(v,r,\theta,\phi)$ where $v$ is null, for definiteness we will not introduce $\epsilon$, but it can be easily incorporated. The line element under the assumptions reads
\begin{equation}
ds^2=-\left(1-\frac{2 r m}{\Sigma}\right)(dv-a \sin^2 \theta d\phi)^2+2 (dv-a \sin^2 \theta d\phi)(dr-a \sin^2 \theta d\phi)+\Sigma (d\theta^2 +\sin^2 \theta d\phi^2), \label{met2}
\end{equation}
where $\Sigma :=r^2+a^2 \cos^2 \theta$ and $m=m(v,r,\theta)$.
Then (\ref{trace}) reduces to
\begin{equation}
 \frac{2}{\Sigma} \frac{\partial^2}{\partial r^2}\left(-rm +2 \alpha \frac{r^2 m^2 \xi}{\Sigma^3} \right)- 4 \Lambda =0,
 \label{quadeq}
\end{equation}
with $\xi:= r^2-3 a^2 \cos^2{\theta}$. As in the spherically symmetric case, (\ref{quadeq}) also gives a quadratic equation for $m$
\begin{equation}
r m-2 \alpha \frac{r^2 m^2 \xi}{\Sigma^3 }+\Lambda \left(r^2 \Sigma-\frac{5 r^4}{6}\right)=p r+q, \label{karmasik}
\end{equation}
where $p=p(v,\theta)$ and $q=q(v,\theta)$ are arbitrary functions of $v$ and $\theta$. The solutions are
\begin{equation}
m(v,r,\theta)= \frac{ \Sigma^3}{ 4 \alpha \xi r } \left ( 1 \pm \sqrt{ 1 - \frac{ 8 \alpha \xi}{ \Sigma^3} \left ( p r + q - \Lambda \left ( \Sigma r^2 - \frac{5}{6} r^4 \right ) \right ) } \right). \label{sol2}
\end{equation}
which is a non-stationary generalization of the one given in \cite{Fernandes} with, also, a nonzero cosmological constant.  Observe that the cosmological constant drastically changes the solution.

Let us now calculate the horizon structure of this solution. A null surface defined by $\mathcal{H}(v,r,\theta)=$ constant  satisfies $g^{\mu \nu}\, \partial_{\mu}\mathcal{H} \partial _{\nu}\mathcal{H}=0$, which, for the metric (\ref{met2}), becomes
\begin{equation}
a^2 \sin^2\theta (\partial_v \mathcal{H})^2+\left(-2 r m+a^2+r^2\right)(\partial_r \mathcal{H})^2 +2 \left(a^2+r^2\right) \partial_r \mathcal{H} \partial_v \mathcal{H}+ (\partial_\theta \mathcal{H})^2=0,
\end{equation}
and $m$ given in (\ref{sol2}) should be inserted in this equation. It is a highly non-trivial PDE. We can make a further assumption for the null horizon's coordinates: that is $\mathcal{H}= r- r_H(v,\theta)=0$, then the horizon equation reduces to
\begin{equation}
a^2 \sin^2\theta (\partial_v r_H)^2+\left(-2 r_H m+a^2+r_H^2\right) -2 \left(a^2+r_H^2\right) \partial_v r_{H}+ (\partial_\theta r_{H})^2=0.
\end{equation}
Let us make some remarks on this equation.  Since  $r_{H}(v,\theta)$ satisfies a first-order nonlinear partial differential equation, it is still very hard to get a closed-form solution, but by linearizing around the stationary solution \footnote{This stationary solution corresponds to the one given in \cite{Fernandes} for the case $\Lambda =0$, otherwise, it is more general than that solution.} we can find approximate solutions. To this end, let $\epsilon$ be a small parameter and expand the non-stationary horizon radius  $r_{H}(v,\theta)$ around the stationary one as
\begin{equation}
r_{H}(v,\theta)= r^{0}_{H}(\theta)+\epsilon r_{1}(v, \theta)+ \mathcal{O}(\epsilon^2),
\end{equation}
where $r^{0}_{H}(\theta)$ is the $v$-independent horizon function that satisfies
\begin{equation}
(\partial_\theta r^0_{H})^2+\left(-2 r^0_H m^0(r,\theta)+a^2+{(r^0_H)}^2\right) =0.
\end{equation}
where  $m^0(r,\theta)$ follows from (\ref{sol2}) with constant $q$ and $p$.
Then $r_{1}(v, \theta)$  satisfies the following linear, but still a partial differential, equation
\begin{equation}
-\left (a^2+(r^{0}_{H})^2\right)\, \partial_{v}\,r_{1}+(\partial_{\theta}\, r^{0}_{H})\, \partial_{\theta}\,r_{1}=(m_{0}+\zeta r^{0}_{H}-r^{0}_{H}) r_{1} \label{eq1},
\end{equation}
where we set $m(v,r,\theta)=m_{0}(v,r)+\epsilon \zeta(r,\theta) r_{1}(v,\theta)$ with $\zeta$ a cumbersome but known function from the expansion (\ref{sol2}). Assuming $\partial_{\theta}\, r^{0}_{H} \ne 0$, we can solve (\ref{eq1}) with the following ansatz:
\begin{equation}
r_{1}(v,\theta)=e^{\rho(\theta) }\,f(v,\theta), \label{eq2}
\end{equation}
which leads to
\begin{equation}
\rho(\theta)=\int^{\theta}_{\theta_0}\, \frac{m_{0}+(\zeta-1) r^{0}_{H}}{\partial_{\theta}r^{0}_{H}}\, d\theta,
\end{equation}
and  $f(v,\theta)= f(\eta)$ is an arbitrary function of $\eta$ which is given as
\begin{equation}
\eta :=v+\int^{\theta}_{\theta_0}\, \frac{a^2+(r^{0}_{H})^2}{\partial_{\theta}\,r^{0}_{H}}\, d\theta.
\end{equation}
Since the linearized solution $r_{1}(v,\theta)$ contains the arbitrary function $f$ depending on $v$ and $\theta$, it may not be bounded for certain values of $v$ and $\theta$. Hence the radiating extension of the rotating black hole, as in the case of the static black studied in Section II, violates the second law of the black hole thermodynamics, and the stability of the horizons is not guaranteed.

Let us study the asymptotic structure of the two curvature invariants of the radiating metric.

\begin{itemize}

\item as $r \rightarrow \infty$, one has

\begin{eqnarray}
&&R= 4 \Lambda+ \frac{30 r^2}{\alpha \sin\theta}+\frac{6 a^2\cos^2\theta}{\alpha  \sin\theta }+\mathcal{O}\left(\frac{1}{r^2}\right), \nonumber \\
&&\mathcal{G}=\frac{112 r^4 }{\alpha ^2 \sin^2 \theta}-\frac{112 r^2 a^2 \cot ^2\theta}{\alpha ^2}-\frac{32 a^4 \cos ^2\theta \cot ^2\theta }{\alpha ^2}-\frac{48  p(v)}{\alpha r \sin\theta }+\mathcal{O}\left(\frac{1}{r^2}\right).
\end{eqnarray}
\item The expressions as $r\rightarrow 0$ are cumbersome. So we assume $q(v)=0$ for the sake of depicting purposes
\begin{eqnarray}
&&R=-\frac{8 \alpha  \tan \theta \sec ^7\theta p(v)^2}{a^8}+\frac{6 a^2 \cos \theta \cot \theta}{\alpha }+4 \Lambda +\mathcal{O}\left(r\right), \nonumber \\
&&\mathcal{G}=\frac{48 \sec ^6\theta p(v)^2}{a^6}-\frac{32 a^4 \cos ^2 \theta \cot ^2 \theta }{\alpha ^2}+O\left(r\right).
\end{eqnarray}
Both of these curvature invariants are finite, unlike the Kerr-black hole case which has a ring-like singularity.
\end{itemize}

\section{Conclusions and Discussions}

Motivated by two recent developments, that is the construction of stationary rotating black hole solutions of the conformal anomaly sourced General Relativity \cite{Fernandes}, and the observation of the amplification of quantum corrections by extremal black holes \cite{Horo}, we have studied here radiating non-rotating and rotating black hole solutions of the $A$-type anomaly sourced  General Relativity with a cosmological constant. Our solutions generalize the rotating solution of \cite{Fernandes} and the spherically symmetric solutions of \cite{Cai1,Cai2} to the non-stationary case as akin to Vaidya's generalization of the Schwarzschild metric, and a similar generalization of the Kerr metric. The metrics we have found are highly non-trivial: we have found the event horizon equations but we could only solve them analytically in the linearized approximation of their stationary counterparts. A numerical investigation of these equations and a proper understanding of the geometric structure of these black holes would be valuable. 

Finally, we would like to point out that the only field equation to determine the metric of spacetime is the trace equation (\ref{an}). It may not be seen as natural or physical to talk about a "solution" for a metric obtained solely from one equation, and it may not be possible to draw some conclusions from such a metric in four dimensions. Hence it is hard to call any metric obtained from the trace equations  (\ref{an}) a "solution". Despite this, we wanted to draw attention to how black hole physics is altered with the trace anomaly. For that reason, we studied a metric that fits our purpose which is the well-known Kerr-Schild metric (\ref{met2}). Indeed, we showed that many important properties of black holes of general relativity may change under the back reaction effects or the conformal anomaly.

\newpage

\section{Appendix A: Energy Momentum Tensor}
The metric in (\ref{met2}) can be written in the Kerr-Schild form
\begin{equation}
g_{\mu \nu}=g^{0}_{\mu \nu}+ \frac{2 m r}{\Sigma}\, \lambda_{\mu}\, \lambda _{\nu}, \label{met3}
\end{equation}
where $\lambda_{\mu}:=(1,0,0,-a \sin^2 \theta)$ is a null vector and  $g^{0}_{\mu \nu}$ is the flat metric withe line element
\begin{equation}
ds^2=-(dv-a \sin^2 \theta d\phi)^2+2 (dv-a \sin^2 \theta d\phi)(dr-a \sin^2 \theta d\phi)+\Sigma (d\theta^2 +\sin^2 \theta d\phi^2).
\end{equation}
Following a similar discussion as in \cite{Fernandes}, one can rewrite the Einstein tensor and hence energy-momentum tensor components as
\begin{equation}\label{enmom}
\langle T_{\mu \nu} \rangle=\lambda_{\mu} \zeta_{\nu}+\lambda_{\nu} \zeta_{\mu}+\rho_{4}\, n_{\mu}\, n_{\nu}+\rho_{6}\, k_{\mu}\, k_{\nu}+\mu\, g_{\mu \nu},
\end{equation}
or more explicitly
\begin{eqnarray}
&& \langle T_{\mu \nu} \rangle=\frac{\rho_{6}}{a^2\, \sin^4 \theta} \lambda_{\mu}\, \lambda_{\nu}+\lambda_{\mu} \left(\zeta_{\nu}-\frac{\rho_{6}}{a^2\, \sin^4 \theta} \,t_{\nu}\right)+\lambda_{\nu} \left(\zeta_{\mu}-\frac{\rho_{6}}{a^2\, \sin^4 \theta} \,t_{\mu}\right)+\rho_{4}\, n_{\mu}\, n_{\nu} \nonumber\\
 &&\quad \quad+\frac{\rho_{6}}{a^2\, \sin^4 \theta}\,t_{\mu}\, t_{\nu} +\mu\, g_{\mu \nu},
\end{eqnarray}
where $\zeta_{\mu} :=\rho_{2}\,n_{\mu}+\rho_{3}\, m_{\mu}+\rho_{5}\, k_{\mu}$ and
\begin{eqnarray}
&&\lambda_{\mu}=(1,0,0,-a \sin^2 \theta), \quad \quad \lambda^{\mu}=(0,1,0,0) \quad \quad
t_{\mu}=(1,0,0,0),   \nonumber \\
&&m_{\mu}=(0,1,0,0), \quad \quad\quad\quad
n_{\mu}=(0,0,1,0),\quad \quad  \quad k_{\mu}=(0,0,0,1).
\end{eqnarray}
One can see that
\begin{equation}
\lambda^{\mu}\, \langle T_{\mu \nu}\rangle =-\frac{2 r^2 \partial_r m}{\Sigma^2}\, \lambda_{\nu}.
\end{equation}
The functions $\rho_{2}, \rho_{3}, \rho_{4}, \rho_{5}, \rho_{6}$ , $\mu$ and the $R$ curvature are given by

\small
\begin{eqnarray}\nonumber
&&\rho_{2}=\frac{1}{\Sigma^2}\, \left(2 a^2 \cos^2 \theta\, m_{\theta}-2 a^2 r \cos \theta \sin \theta m_{v}+r^2 \Sigma m_{r \theta}-\Sigma m_{\theta} \right),\\
&&\rho_{3}= \frac{1}{\sin \theta \, \Sigma^2\, (2 m r-\Sigma)} \left(2 a^2 \Sigma \, \sin^3 \theta \, m_{vr}+a^2 r \Sigma \sin^3 \theta \, m_{rr} \right. \nonumber \\
&&+2 a^2 \sin^3 \theta \,(-2 a^2 \sin^2 \theta+2a^2 -\Sigma)\, m_{r} \nonumber \\
&&+ r\, \cos \theta \, (4 a^2 \sin^2 \theta+\Sigma)\, m_{\theta}+ a^2 r \sin^3 \theta \, \Sigma\, m_{vv}+ 2 \sin \theta (2 a^4 \, \cos^2 \theta\, \sin^2 \theta+ \nonumber \\
&&\left. 2a^2 \Sigma\, \cos^2 \theta-a^2 \Sigma- \Sigma^2)\,m_{v}+r \sin \theta\, \Sigma\, m_{\theta \theta} \right),\nonumber\\
&&\rho_{4}= \frac{1}{\sin \theta \, \Sigma\, (2 m r-\Sigma)} \left(2a^2r \sin^3 \theta \, m_{vr}+\sin \theta\, \Sigma\, (-2a^2 m \sin^2 \theta+a^2 r \sin^2 \theta+2 a^2 m-2m \Sigma +r \Sigma)\, m_{rr} \right.\nonumber \\
&&+2 \sin \theta\,(2 a^4 \sin^2 \theta \cos^2 \theta+4 a^2 mr \sin^2 \theta-3 a^2 \Sigma \, \sin^2 \theta-4 a^2 mr+2a^2 \Sigma+2mr \Sigma-\Sigma^2)\, m_{r}+ \nonumber \\
&& r \cos \theta\,(4a^2 \cos^2 \theta+\Sigma)\, m_{\theta}+ a^2 r \Sigma\, \sin^3 \theta\, m_{vv}+2 \sin \theta\,(2 a^4 \cos^2 \theta \sin^2 \theta-2 a^2 \Sigma \, \sin^2 \theta+a^2 \Sigma-\Sigma^2)\, m_{v} \nonumber \\
&& \left.+r \Sigma \, \sin \theta\,\, m_{\theta \theta}
\right),\nonumber\\
&&\rho_{5}=\frac{1}{\Sigma^2\, (2 m r-\Sigma)} \left(a \,  \sin^2 \theta \, \Sigma \,(2 a^2 m \, \cos^2 \theta+2 a^2 r \sin^2 \theta+ r \Sigma-2 m \Sigma)\, m_{vr} \right. \nonumber\\
&&+a \sin^2 \theta \, \Sigma\, (2 a^2 m \cos^2 \theta +a^2 r \sin^2 \theta  -2m \Sigma +r \Sigma)\, m_{rr}+2 a \sin^2 \theta \,(2 a^4 \cos^2 \theta \sin^2 \theta-4 a^2 m r \cos^2 \theta + \nonumber \\
&&3 a^2 \Sigma \, \cos^2 \theta-a^2 \Sigma +2mr \, \Sigma-\Sigma^2)\, m_{r}+a \cos \theta \, \sin\theta \,(4 a^2 m \cos^2 \theta+4 a^2 r \sin^2 \theta-4m \Sigma +3 r \Sigma)\, m_{\theta} \nonumber \\
&&+a^3 r \Sigma \sin^4 \theta \, m_{vv}+a \sin^2 \theta\, (4 a^4 \cos^2 \theta\, \sin^2 \theta-4a^2 mr \cos^2 \theta+6 a^2 \Sigma \cos^2 \theta-2 a^2 \Sigma+2mr \Sigma-3 \Sigma^2)\, m_{v} \nonumber\\
&&\left.+a r \Sigma \sin^2 \theta\, m_{\theta \theta} \right), \nonumber\\
&&\rho_{6}=\frac{1}{ \Sigma\, (2 m r-\Sigma)} \left(-2 a^2 r \Sigma \sin^4 \theta \, m_{vr}+\Sigma \, \sin^2 \theta\,(2 a^2 m \sin^2 \theta-a^2 r \sin^2 \theta-2 a^2 m+2m \Sigma-r \Sigma)\,m_{rr} \nonumber \right.\\
&&+2 \sin^2 \theta\, (-2 a^2 \sin^2 \theta \, \cos^2 \theta+ 4 a^2 m r \cos^2 \theta+3 a^2 \Sigma \sin^2 \theta-2 a^2 \Sigma-2mr \Sigma+\Sigma^2)\, m_{r}\nonumber \\
&&+r \sin \theta \cos \theta\,(4 a^2 \sin^2 \theta+\Sigma)\, m_{\theta}-a^2 r \Sigma \sin^4 \theta \,m_{vv} \nonumber \\
&&+2 \sin^2 \theta\, (-2 a^4 \sin^2 \theta \cos^2 \theta+2 a^2 \Sigma \sin^2 \theta-a^2 \Sigma+\Sigma^2)\, m_{v} \nonumber \\
&&\left.-r \Sigma \sin^2 \theta \, m_{\theta \theta} \right), \nonumber \\
&&\mu=\frac{1}{\sin \theta \, \Sigma^2\, (2 m r-\Sigma)} \left(-2 a^2 r \, \Sigma \sin^3 \theta \, m_{vr}-a^2 r\, \sin^3 \theta\, \Sigma \, m_{rr}-r \cos \theta\, (\Sigma+4 a^2 \, \sin^2 \theta)\, m_{\theta} \right.\nonumber\\
&& 2 \sin \theta\, (-2 a^4 \cos^2 \theta \, \sin^2 \theta+2 a^2 mr \cos^2 \theta- 2 a^2 \, \cos^2 \theta\, \Sigma +a^2 \, \Sigma- 2mr \Sigma+\Sigma^2)\,m_{r} -a^2 r \, \sin^3 \theta \, \Sigma\, m_{vv}\nonumber\\
&& \left.+2 \sin \theta \,(-2 a^4 \cos^2 \theta \, \sin^2 \theta- 2 a^2 \, \Sigma\, \cos^2 \theta+a^2 \, \Sigma+\Sigma^2)\,m_{v}-r \, \sin \theta\, \Sigma \, m_{\theta \theta}   \right),\nonumber\\
&&R=\frac{1}{ \Sigma\, (-2 m r+\Sigma)} \left(2(2 a^2 m \cos^2 \theta-2m \Sigma+r \Sigma)\,m_{rr}+4 (-2mr+\Sigma) \right),
\end{eqnarray}
\normalsize
with the identy $\rho_{6}=\sin^2  \theta\, \rho_{4}+\sin^2 \theta\, \Sigma \, (R+2 \rho_{3}+4 \mu)$.
Here $m_r := \partial_r m$ {\it etc.}. 

\vspace{1.5 cm}
\noindent
{\bf Energy Conditions}: Let $U^{\mu}$ be any timelike or a null vector in the spacetime geometry, then it is not possible to say anything about the sign of the term $U^{\mu} U^{\nu}\, \langle T_{\mu \nu} \rangle$, because
\begin{equation}\label{enercond}
U^{\mu} U^{\nu}\, \langle T_{\mu \nu} \rangle=2 (U \cdot \lambda)(U \cdot \zeta)+\rho_{4} (U \cdot n)^2+ \rho_{6} (U \cdot k)^2+\mu U^2,
\end{equation}
where $U \cdot \lambda=U^{\mu}\, \lambda_{\mu}$, $U \cdot k=U^{\mu}\, k_{\mu}$, $U \cdot n=U^{\mu}\, n_{\mu}$ ,  $U \cdot \zeta=U^{\mu}\, \zeta_{\mu}$ and $U^2=U^{\mu} U_{\mu}$. For an arbitrary timelike or null vector  $U^\mu$, the right-hand side of (\ref{enercond}) can take any sign. Hence we we cannot deduce any of the energy conditions for the energy momentum tensor given in (\ref{enmom}). In any case, we know that all the known energy conditions can be violated by quantum fields.


\begin{thebibliography}{99}

\bibitem{Hawking}
S.~W.~Hawking,
Particle Creation by Black Holes,
Commun. Math. Phys. \textbf{43}, 199-220 (1975),
[erratum: Commun. Math. Phys. \textbf{46}, 206 (1976)].

\bibitem{Fernandes}
P.~G.~S.~Fernandes, Rotating black holes in semiclassical gravity, Phys. Rev. D {\bf 108}, L061502 (2023).


\bibitem{Cai1}
R.~G.~Cai, L.~M.~Cao and N.~Ohta,
Black Holes in Gravity with Conformal Anomaly and Logarithmic Term in Black Hole Entropy,
JHEP \textbf{04}, 082 (2010).

\bibitem{Cai2}
R.~G.~Cai,
Thermodynamics of Conformal Anomaly Corrected Black Holes in AdS Space,
Phys. Lett. B \textbf{733}, 183-189 (2014).

\bibitem{Duff}
M.~J.~Duff,
Twenty years of the Weyl anomaly,
Class. Quant. Grav. \textbf{11}, 1387-1404 (1994).

\bibitem{Deser1}
S.~Deser and A.~Schwimmer,
Geometric classification of conformal anomalies in arbitrary dimensions,
Phys. Lett. B \textbf{309}, 279-284 (1993).

\bibitem{Mottola}
E.~Mottola and R.~Vaulin,
`Macroscopic Effects of the Quantum Trace Anomaly,
Phys. Rev. D \textbf{74}, 064004 (2006).

\bibitem{Horo}
G.~T.~Horowitz, M.~Kolanowski, G.~N.~Remmen and J.~E.~Santos,
Extremal Kerr Black Holes as Amplifiers of New Physics,
Phys. Rev. Lett. \textbf{131}, no.9, 091402 (2023).

\bibitem{Vaidya}
P.~C.~Vaidya, The external field of a radiating star in general relativity, Curr. Sci. {\bf 12} (6) (1943) 183.
[Republished as Gen. Rel. Grav. \textbf{31}, 119-120 (1999). ]

\bibitem{Ho}
P.~M.~Ho, H.~Kawai, H.~Liao and Y.~Yokokura,
4D Weyl Anomaly and Diversity of the Interior Structure of Quantum Black Hole,
[arXiv:2307.08569 [hep-th]].


\bibitem{geometry}
A.~Coudray and J.~P.~Nicolas,
Geometry of Vaidya spacetimes,
Gen. Rel. Grav. \textbf{53}, no.8, 73 (2021).

\bibitem{Fodor}
G.~Fodor, K.~Nakamura, Y.~Oshiro and A.~Tomimatsu,
Surface gravity in dynamical spherically symmetric space-times,
Phys. Rev. D \textbf{54}, 3882-3891 (1996).

\bibitem{Nielsen}
A.~B.~Nielsen and J.~H.~Yoon,
Dynamical surface gravity,
Class. Quant. Grav. \textbf{25}, 085010 (2008).

\end{thebibliography}
\end{document}